\documentclass[11pt]{article}
\usepackage[english]{babel}

\usepackage[margin=1in]{geometry}

\usepackage[english]{babel}
\usepackage[utf8]{inputenc}

\usepackage[T1]{fontenc}
\usepackage{kpfonts}
\usepackage{libertine}
\usepackage[scaled=0.85]{beramono}

\usepackage{amssymb}
\usepackage{amsfonts}
\usepackage{amsmath}
\usepackage{bbm}
\usepackage{graphicx}
\usepackage[colorlinks=true, allcolors=blue]{hyperref}
\usepackage{algorithm}
\usepackage{algpseudocodex}
\usepackage{amsthm}
\usepackage{thmtools}
\usepackage{thm-restate}
\usepackage{mathtools}
\usepackage{verbatim}

\usepackage{mdframed}

\usepackage{wrapfig}
\newcommand{\erclogowrapped}[1]{%
\setlength\intextsep{0pt}%
\begin{wrapfigure}[3]{r}{#1*\real{1.1}}%
\includegraphics[width=#1]{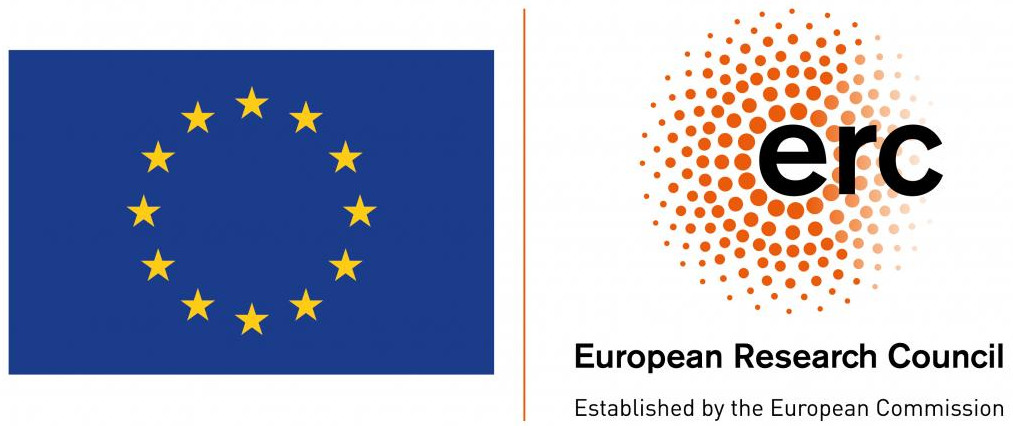}%
\end{wrapfigure}%
}

\usepackage{blkarray, bigstrut}
\makeatletter
\DeclareRobustCommand
  \myvdots{\vbox{\baselineskip4\p@ \lineskiplimit\z@
    \hbox{.}\hbox{.}\hbox{.}}}
\makeatother

\makeatletter
\def\blfootnote{\xdef\@thefnmark{}\@footnotetext}
\makeatother

\theoremstyle{plain}
\declaretheorem{theorem}

\declaretheorem[sibling=theorem]{lemma}
\declaretheorem[sibling=theorem]{corollary}

\theoremstyle{definition}
\declaretheorem[sibling=theorem]{Definition}

\declaretheorem[sibling=theorem]{Remark}

\newcommand{\R}{\mathbb{R}}

\newcommand{\ones}{\mathbf{1}}
\newcommand{\zeros}{\mathbf{0}}
\newcommand{\OPT}{\mathrm{OPT}}
\newcommand{\ALG}{\mathrm{ALG}}
\DeclareMathOperator{\cost}{cost}
\DeclareMathOperator{\len}{len}
\DeclareMathOperator{\poly}{poly}

\renewcommand{\ge}{\geqslant}
\renewcommand{\le}{\leqslant}
\renewcommand{\geq}{\geqslant}

\title{Online Min-Max Paging}
\author{
Ashish Chiplunkar\\
Indian Institute of Technology Delhi\\
\texttt{ashishc@iitd.ac.in}
\thanks{Ashish Chiplunkar is partially supported by the Pankaj Gupta New Faculty Fellowship.}
\and 
Monika Henzinger\\
Department of Computer Science\\
University of Vienna\\
\texttt{monika.henzinger@univie.ac.at}
\and 
Sagar Sudhir Kale\\
Department of Computer Science\\
University of Vienna\\
\texttt{sagar.kale@univie.ac.at}
\and 
Maximilian Vötsch\\
Department of Computer Science,\\UniVie Doctoral School Computer Science DoCS\\
University of Vienna\\
\texttt{maximilian.voetsch@univie.ac.at}
}
\date{}
\begin{document}
\maketitle

\pagenumbering{gobble}
\thispagestyle{empty}
\begin{abstract}
Motivated by fairness requirements in communication networks, 
we introduce a natural variant of the online paging problem, called \textit{min-max} paging, where the objective is to minimize the maximum number of faults on any page. While the classical paging problem, whose objective is to minimize the total number of faults, admits $k$-competitive deterministic and $O(\log k)$-competitive randomized algorithms, we show that min-max paging does not admit a $c(k)$-competitive algorithm for any function $c$. Specifically, we prove that the randomized competitive ratio of min-max paging is $\Omega(\log(n))$ and its deterministic competitive ratio is $\Omega(k\log(n)/\log(k))$, where $n$ is the total number of pages ever requested.

We design a fractional algorithm for paging with a more general objective -- minimize the value of an $n$-variate differentiable convex function applied to the vector of the number of faults on each page. This gives an $O(\log(n)\log(k))$-competitive fractional algorithm for min-max paging. We show how to round such a fractional algorithm with at most a $k$ factor loss in the competitive ratio, resulting in a deterministic $O(k\log(n)\log(k))$-competitive algorithm for min-max paging. This matches our lower bound modulo a $\poly(\log(k))$ factor.
We also give a randomized rounding algorithm that results in a $O(\log^2 n \log k)$-competitive algorithm.

\blfootnote{\erclogowrapped{5\baselineskip} \emph{M. Henzinger and M. Vötsch:} This project has received funding from the European Research Council (ERC) under the European Union's Horizon 2020 research and innovation programme (Grant agreement No. 101019564 ``The Design of Modern Fully Dynamic Data Structures (MoDynStruct)'' and from the Austrian Science Fund (FWF) project ``Fast Algorithms for a Reactive Network Layer (ReactNet)'', P~33775-N, with additional funding from the \textit{netidee SCIENCE Stiftung}, 2020--2024.}
\end{abstract}

\pagenumbering{arabic}
\section{Introduction}
\label{sec:intro}


Paging is a decades-old, classical computer science problem. Suppose a computer process working on $n$ \emph{pages} of data has access to two levels of memory: a fast memory, called the \emph{cache}, that can hold a small amount $k$ of pages, and a slow memory containing all $n$ pages. Typically, $k$ is much smaller than $n$, and initially, all pages are in slow memory. Whenever the process accesses a page, it is read from the cache; if it is not already in the cache, a \emph{page fault} occurs, and the page must be brought into the cache, which possibly necessitates \emph{evicting} another page from the cache to make room. This is called \emph{serving} the request. In the \emph{online} setting, each request must be served before the algorithm sees the subsequent request. The goal is to minimize the \emph{total number of page faults} incurred while serving a sequence of requests.

The paging problem has found new applications in communication networks, where caching is ubiquitous and is used to minimize energy usage, communication latency, and network traffic.  
Consider, for example, TCP connections that are kept alive on a router~\cite{cohen1999connection} or optical links in reconfigurable data center topologies~\cite{foerster2019survey, bienkowski2021online}. Every user application prefers to have an active connection, as re-establishing a TCP connection or link takes time and slows down communication or computation. Another example is content on web pages that is cached in a content delivery network, such as Akamai. In practice, the cache servers in these networks rely on dynamic, eviction-based algorithms for managing cache contents that solve the so-called content placement problem~\cite{thomdapu2021dynamic}. Web pages in the cache have a clear advantage as they can be served faster to the user than web pages that must be re-fetched from the server. 
Ideally, all applications (of the same priority) and all web pages should be treated equally.
This motivates us to propose the study of a \emph{fair} variant of paging, which we call \textit{min-max paging}. Its goal is to minimize the number of page faults on \emph{any} page, i.e.~to minimize the \emph{maximum number of page faults of any single page}.


In the \emph{online} setting, the page requests are revealed one by one without  knowledge of the future, so the description of how to serve each request must depend only on the request sequence thus far and the current cache contents. Naturally, for many problems, an online algorithm cannot output an optimal solution to a given instance -- something an {\em offline} algorithm having access to the entire input can produce. The sub-optimality of an online algorithm is usually measured using competitive analysis. Informally, we say that an online algorithm has a competitive ratio of $c$ if, on every problem instance, it produces a solution with (expected) cost at most $c$ times the cost of an optimal solution.
For the classic online paging problem the competitive ratio has been well-studied: It is $O(k)$ for deterministic algorithms~\cite{SleatorT85,karlinMRS86}
 and $H_k \approx 0.577 + \ln(k)$ for randomized algorithms~\cite{FiatKLMSY_JAlg91,10.1007/BF01759073}.
To the best of our knowledge, the min-max paging problem has not been studied before. While an efficient offline algorithm for the classical paging problem is known, we neither have an efficient offline algorithm for min-max paging nor a proof of NP-hardness. In this paper, we focus on the min-max paging problem in the {\em online} setting and give both upper and lower bounds on its competitive ratio. 

\paragraph{Our results} We first propose an algorithm for the \textit{fractional paging problem} with objective function $f$, where pages can be held in the cache fractionally, subject to having a total volume of at most $k$ pages at all times. The objective function $f$ is an arbitrary function from an appropriately defined subclass of convex functions applied to the \textit{fault vector}. Here, the fault vector refers to the $n$-dimensional vector of the number of faults incurred on each page, where $n$ is the number of pages. We use the theory of convex programming and properties of $f$ to analyze our algorithm and establish the following bound.
\begin{theorem}[Stated formally as Theorem~\ref{thm:frac}]
For the fractional paging problem with objective function $f$, there exists a $(2q\log(k+1))^q$-competitive algorithm, provided $f$ grows no faster than a degree-$q$ polynomial.
\end{theorem}
In particular, when instantiating $f$ to be the $q$'th power of the $q$-norm, we get the following bound:
\begin{restatable}[Stated formally as Theorem~\ref{qpaging}]{theorem}{qpaging}
For the fractional paging problem with the objective of minimizing the $\ell_q$-norm of the fault-vector, there exists a $2q\log(k+1)$-competitive algorithm.
\end{restatable}
Note that the above theorem does not give a sensible result for the $\ell_\infty$-norm, which is the objective function we are interested in. However, using the fact that the $\ell_\infty$-norm of an $n$-dimensional vector is well-approximated by its $\ell_{\log(n)}$-norm, we get the following result.
\begin{theorem}[Stated formally as Theorem~\ref{thm_frac_ub}]
For the fractional paging problem with the objective of minimizing the $\ell_\infty$-norm of the fault-vector (a.k.a. fractional min-max paging), there exists an $O(\log(n)\log(k))$-competitive algorithm.
\end{theorem}
Next, we propose two approaches for rounding solutions of fractional min-max paging algorithms online and obtain the following two results. Note that the bound of the latter result is better than the former in the $k=\omega(\log n)$ regime, and it also rules out a lower bound linear in $k$ for randomized algorithms.
\begin{theorem}[Stated formally as Corollary~\ref{detround}]
  There exists an $O(k \log(k) \log(n))$-competitive {\em deterministic} algorithm for min-max paging.
\end{theorem}
\begin{restatable}[]{theorem}{randomizedfinal}
  There exists an $O(\log^{2}n \log k)$-competitive {\em randomized} integral algorithm for min-max paging.
\end{restatable}
We complement the above upper bounds by the following impossibility results.
\begin{theorem}[Stated formally as Theorem~\ref{thm:detlb}]
  Every deterministic algorithm for min-max paging is $\Omega(k \linebreak[2] \log(n)/ \log(k))$-competitive.
\end{theorem}
\begin{theorem}[Stated formally as Theorem~\ref{thm_lb}]
  Every algorithm for min-max paging is $\Omega(\log(n))$-competitive.
\end{theorem}
Note that we only have a $O(\log^2k)$ discrepancy between our deterministic bounds, i.e., the bounds are tight up to a polylogarithmic in $k$ factor. Moreover, our lower bounds show that min-max paging is fundamentally more difficult than classical paging and its several generalizations (see Section~\ref{sec:related-work}), which admit competitive ratios independent of $n$, the total number of pages.


We now present some intuition why algorithms for the classical paging problem and a simple algorithm for min-max paging fail to achieve anything better than a trivial competitive ratio for min-max paging. Algorithms for the classical paging problem are oblivious to the number of faults a single page has incurred while processing the sequence $\sigma$ up to a given point in time $t$. Consider the Least Recently Used (LRU) algorithm, which evicts the page whose last request was before the requests to other pages in the cache. Let $p_0 \in P = \{p_0, p_1, \dots, p_n\}$ and assume that $n = |P| - 1 = mk$ is a large multiple of $k$. The sequence $\sigma = (p_0, p_1,p_2, \dots, p_k, p_0, p_{k+1}, p_{k+2}, \dots, p_{2k}, p_0, p_{2k+1}, \dots, p_{mk}, p_0)$ will cause the LRU algorithm to fault $m + 1$ times on page $p_0$, while the optimal algorithm faults exactly once per page. We pair each request to $p_0$ with requests to a set of $k$ pages. After processing these $k$ requests, LRU will have ejected $p_0$, so the next request to $p_0$ will result in a page fault, yielding in a competitive ratio of $\Omega(n/k)$.

Another obvious strategy is to greedily keep the $k$ pages which have incurred the most faults thus far in the cache. In this case, there also exists a request sequence for which the strategy is no better than $\frac{n}{k}$-competitive. For simplicity's sake, let us assume that $k=2$. Then the request sequence is constructed as follows:
(1) Request $p_1, p_2, p_3$ in this order $N$ times, where $N$ is a parameter.
(2) Request $p_4, p_5$ until the algorithm includes \emph{both} of them into the cache.
(3) Request $p_4, p_5, p_6$ in this order $N$ times.
(4) Repeat steps 2 and 3 with pages $p_7$, $p_8$, and $p_9$ next, then with pages $p_{10}$, $p_{11}$, and $p_{12}$, and so on.

After step 3, the greedy algorithm will hold two pages of cost $(r+1)N$, where $r$ is the number of times we have repeated steps $2$ and $3$. In step 2, we request a set of new pages, and the greedy algorithm will fault on them until they reach cost $(r+1)N$. During step $3$, the algorithm will fault $N$ times on each page, making it so that it has cost $(r+2)N$ on the pages introduced in step $2$. Meanwhile, the cost of the optimal offline algorithm is no greater than $N$, obtained by immediately adding the pages of step 2 to the cache.

\paragraph{Our techniques}
Our lower bound of $\Omega(k \log(n)/ \log(k))$ is established by generalizing the above construction, using the following approach: The adversary fixes a sequence of requests over a set of $n$ pages, which can be served while keeping the number of faults on any page small. The core idea is to successively reduce the set of pages that we request in the future in such a way that the algorithm \emph{cannot predict which pages will stop being requested.} A clairvoyant adversary processes the sequence so that she initially incurs a small number of faults on pages that will be requested many times in the future. This causes the adversary to have roughly uniform cost over all pages, while the algorithm has one page on which it has faulted many times.

To design an online algorithm 
one could try to use  standard techniques to transform a max-based objective function into a linear program and solve the corresponding linear program online. However, this  does not work as all known online algorithms for linear programs only work with {\em exclusively packing or exclusively covering constraints} and can not handle a mix of constraints, except for~\cite{azar2013online}, which cannot handle {\em box constraints,} i.e., an upper bound on the variables as required for paging.

Thus, to solve the online min-max paging problem, we solve a more general problem: We give a $O((q\log k)^q)$-competitive  algorithm for a \emph{fractional  paging problem, which minimizes a convex, differentiable function with $q$-bounded growth} and \emph{an upper bound constraint (i.e., a box constraint) on each variable.} A function with $q$-bounded growth behaves like a polynomial function of degree $q$.
To the best of our knowledge, this problem has not been studied before, and no non-trivial online algorithm is known.

As our cost function is not linear, the combinatorial technique of potential functions used for server problems with linear cost functions breaks down. Informally, a potential function captures the advantage accumulated by the adversary at any time, which she can use to make the algorithm ``pay'' more than herself in the future. The potential function is a function on the \textit{state space} of the problem, where the state of the algorithm, at any time, fully determines its future behavior. The state space is usually a small set when the objective is linear. On the contrary, in the case of min-max paging, a state must capture the vector of faults accumulated on each page and its current cache, and there can be multiple fault vectors for the same current cache, which makes the state space blow up with every request, thus, making the use of potential functions challenging, messy, and inelegant.

Instead, we build on the work of Azar~et~al.~\cite{azarOnlineAlgorithms2016}, which minimizes a convex cost function with linear constraints of \emph{row sparsity} $\rho$. Their approach requires the variables $x_{p,j}$ to be unbounded, and for $q$-bounded growth functions, it gives an $O((q\log\rho)^q)$-competitive algorithm. 

We also draw on ideas from Bansal et al.~\cite{bansalbn_12}, which  studied the weighted paging problem with linear cost functions.
They first compute a fractional solution using a primal-dual approach and then show how to round it.
As they have a linear cost function, they can show that the rate of increase of the primal, i.e., the fractional algorithm's cost, is proportional to the rate of increase of the dual.
In our setting, the cost function is not linear, and we have to use the theory of duality of convex programs and conjugate duals. To do so, we extend their analysis to the convex program setting, which requires solving various technical hurdles. This results in a $(2q \log (k+1))^q$-competitive algorithm for fractional paging with any  convex, differentiable function with $q$-bounded growth and box constraints.
Furthermore, for norm-objective functions, more specifically for $q$-norms, we achieve a  competitive ratio of $2q \log k$.
Since the cost function of min-max paging is the $\ell_\infty$-norm of the vector of page-wise costs, we approximate it by $\ell_{\log n}$-norm, resulting in a $2e \log n \log(k+1)$-competitive algorithm
for fractional min-max paging.

We round our solution deterministically using for every page $p$ a threshold  for $x_p$of $1-1/k$, resulting in the upper bound of $O(k \log n \log k)$.
It might be tempting to apply the randomized rounding algorithm of~\cite{bansalbn_12} directly, but it does not apply as it crucially uses the fact that the cost of the algorithm is the sum of the fractional values of all pages.
Instead, we adapt the scheme of~\cite{bansal2021efficient} from the weighted paging setting to the  min-max setting. Specifically, this requires to ``charge'' the cost of each rounding step to each individual page, as opposed to the sum of the changes in the fractional solution over all pages. This charging to individual pages has not been done before in online paging and might be interesting in other settings.


In Section 2, we give all definitions.  In Section 3, we show our lower bounds,
in Section 4, we present and analyze our algorithm for paging with convex objective functions.
In Section 5 we round the fractional algorithm to obtain an $O(k \log(n) \log(k))$-competitive deterministic and $O(\log^2 (n) \log(k))$-competitive randomized algorithms for (integral) min-max paging. All omitted proofs  are given in the appendix.



\section{Preliminaries}


The problems in this paper are studied in the online setting, where an adversary fixes a request sequence $\sigma$ ahead of time, and the requests in this sequence are presented to an algorithm one by one. When the algorithm receives a new request from the sequence, it can only use its knowledge of the requests seen thus far to make a decision. In particular, the algorithm does not have any knowledge of future requests.

In this setting, we use \emph{competitive analysis} \cite{SleatorT85} to measure the quality of an algorithm. 
In competitive analysis, we study the \emph{competitive ratio} of an online algorithm, which compares the worst-case ratio between the cost of the algorithm and the cost of an optimum offline solution over all possible $\sigma$.

More formally, for a deterministic algorithm $\ALG$, the competitive ratio of $\ALG$ is the smallest $c \in \mathbb{R}$, such that for all instances $\sigma$ of an online minimization problem, we have
\[
  \ALG(\sigma) \le c \cdot \OPT(\sigma) + d,
\]
where $\ALG(\sigma)$ is the cost of the algorithm, $\OPT(\sigma)$ is the cost of the optimum offline solution, and $d$ is some constant independent of $\sigma$. We will call an algorithm fulfilling the above definition a $c$-competitive algorithm. If $\ALG$ is a randomized algorithm, then the competitive ratio is defined as the smallest $c \in \mathbb{R}$ such that
\[
  \mathbb{E}[\ALG(\sigma)] \le c \cdot \OPT(\sigma) + d.
\]



We study a variant of the paging problem called \emph{min-max paging}. In any paging problem the request sequence $\sigma$ is made up of requests to a set of pages $P = \{p_{1}, p_{2}, \dots, p_{n}\}$ of size $n$. We will assume that $\sigma$ is of finite length, denoted by $T$. The algorithm is given a cache $C$ of size $k$, which always is a subset of $P$ and is empty when the algorithm begins processing $\sigma$.

When page $p$ is requested during round $t$, we must add $p$ to the cache $C$ if it is not already contained in $C$. If adding the page causes $C$ to be of size $k+1$, we must evict a page other than $p$ from the cache before we are allowed to process the next request. The situation where a request to page $p$ arrives while $p$ is not in $C$ is called a \emph{page fault}.

Whenever a page fault occurs, we incur some cost. The objective of the classical paging problem is to minimize the total number of page faults. In the case of min-max paging, the objective is to minimize the maximum number of page faults occurring for any page. More precisely, if we let $x_{p,j}$ be a zero-one variable, which denotes that a page fault occurs upon the $j$-th request to page $p$, then we seek to minimize
\[
  \max_{{p \in P}} \sum_{j} x_{p,j},
\]
where the summation is over all requests to $p$. We can think of this as minimizing the $\ell_{\infty}$-norm of the vector
$\vec{c}(\sigma) = (\sum_{j} x_{p_{1},j} , \dots, \sum_{j} x_{p_{n},j} )^\top$,
whereas the classical paging problem is equivalent to minimizing the $\ell_{1}$-norm of $\vec{c}(\sigma)$.


In Section 4 we solve a fractional version of the paging problem for convex objective functions $f(x)$, where $x$ is the vector consisting of the variables $x_{p,j}$, under the assumption that $f(x)$ is well behaved. Of particular interest is the case where $f(x) = \|\vec{c}(\sigma)\|_{q}$, i.e. the $\ell_{q}$-norm. We refer to this case as \emph{$q$-paging}. For details refer to Section 4. 

\begin{Remark}
  Paging problems are studied in the eviction cost model, where fetching a page incurs no cost, and the algorithm pays for evicting a page, and in the fetching cost model, where evicting a page comes without an associated cost, and the algorithm pays for fetching a page. For min-max paging, the cost of these models differs by at most $1$. 
  Said difference occurs on the set of pages contained in the cache at time $T$ that the algorithm does not have to evict anymore.

  Because of this equivalence, we use both models interchangeably in this paper. The lower bounds of Section 3 use the fetching cost model, and the upper bounds of Section 4 use the eviction cost model, as the choice of the respective model simplifies the proofs.
\end{Remark}

\section{Lower Bounds}

\label{sec:lb}
We show a deterministic lower bound of $\Omega(k (\log n)/\log k)$ and a randomized lower bound of $\Omega(\log n)$ (for $k = 2$) on the competitive ratio for min-max paging.  Our lower bounds are based on a simple construction that is cleanly demonstrated with $k = 2$ and can be generalized for $k \ge 2$.
The interested reader will find complete proof for the deterministic lower bound in the appendix in Section~\ref{app:lb}.

\begin{restatable}{@theorem}{detlb}
\label{thm:detlb}
Any deterministic algorithm for min-max paging with cache size $k$ is at least $\frac{k-1}{2} \log_{k+1} n$-competitive, where $n$ is the number of pages.
\end{restatable}

\begin{theorem}\label{thm_lb}
  The randomized competitive ratio of min-max paging is $\Omega(\log n)$, where $n$ is the number of pages when the cache size is $k=2$.
\end{theorem}

By Yao's principle, it suffices to exhibit a probability distribution on input instances, forcing every deterministic online algorithm to perform a factor $\Omega(\log n)$ worse in expectation than the optimum cost. Let $n=3^m$ for some large integer $m$. Our adversarial strategy takes a parameter $N\gg n$ and is defined as follows.
\begin{algorithm}
  \caption{An adversarial strategy for min-max paging}\label{adversary}
  \begin{algorithmic}[1]
    \State Let $L_m=\{p^m_0,\ldots,p^m_{n-1}\}$ be a set of $n=3^m$ pages.
    \For{$\ell$ $=$ $m$ to $1$} \State $L_{\ell-1}\gets\emptyset$.  \For{$i$ $=$
      $0$ to $3^{\ell-1}-1$} \State Give $N$ requests to each of
    $p^{\ell}_{3i},p^{\ell}_{3i+1},p^{\ell}_{3i+2}$ in a round-robin manner.
    \State $p^{\ell-1}_i$ $\gets$ a uniformly random page from
    $\{p^{\ell}_{3i},p^{\ell}_{3i+1},p^{\ell}_{3i+2}\}$.  \State Add
    $p^{\ell-1}_i$ to $L_{\ell-1}$.  \EndFor \EndFor
  \end{algorithmic}
\end{algorithm}

We call each iteration of the outer for-loop a \textit{layer} and each of the inner for-loop a \textit{phase}. We number the layers $m,m-1,\ldots,1$.


\begin{lemma}\label{lem_lb_adv}
  The adversary can serve all requests while faulting at most $m+N$ times on every page with probability one.
\end{lemma}
\begin{proof}
  Consider an arbitrary phase of an arbitrary layer $\ell$. Let $q$ be the page added to $L_{\ell-1}$ at the end of the phase, and let $q_1,q_2$ be the other two pages requested in the phase. On the first request to $q$, the adversary will add $q$ to its cache and keep it there until the end of the phase. It uses the remaining cache slot to serve all requests to $q_1$ and $q_2$. Thus, the adversary faults only once on $q$ and $N$ times on $q_1$ and $q_2$ each.

  Consider an arbitrary page $p$. In all phases where $p$ is requested except the last one, the adversary faults only once on $p$ ($p$ is the page $q$ in the above argument). In the last phase, the adversary faults $N$ times on $p$. Every layer contains at most one phase in which $p$ is requested. Since the number of layers is $m$, the algorithm faults at most $m+N$ times on $p$.
\end{proof}

To analyze the algorithm's performance, let the random variable $X^{\ell}_i$ be the number of the algorithm's faults on the randomly chosen page $p^{\ell}_i$ at the beginning of layer $\ell$.

\begin{lemma}\label{lem_lb_alg}
  For every layer $\ell$ and every $i\in\{0,\ldots,3^{\ell}-1\}$ we have $\mathbb{E}[X^{\ell}_i]\geq(m-\ell)\cdot N/2$.
\end{lemma}

\begin{proof}
  We prove the claim by reverse induction on $\ell$. Recall the numbering of phases and observe that $X^m_i=0$ for all $i\in\{0,\ldots,n-1\}$. Thus, the claim is true for $\ell=m$. Assuming as induction hypothesis that for every $i\in\{0,\ldots,3^{\ell}-1\}$ we have $\mathbb{E}[X^{\ell}_i]\geq(m-\ell)\cdot N/2$, we prove that for every $j\in\{0,\ldots,3^{\ell-1}-1\}$ we have $\mathbb{E}[X^{\ell-1}_j]\geq(m-\ell+1)\cdot N/2$.

  In any phase, since the cache size is $2$ and three pages are requested in a round-robin manner $N$ times each, the total number of faults is at least $3N/2$. This is evident if we consider the behavior of the optimal algorithm for (usual) paging that always evicts the page needed farthest in the future. Consider the $j$'th phase of layer $\ell$, and recall that $p^{\ell-1}_j$ is defined at the end of this phase. The total number of faults in this phase is at least $3N/2$, and these faults are distributed over the three pages, $p^{\ell}_{3j},p^{\ell}_{3j+1},p^{\ell}_{3j+2}$. Since $p^{\ell-1}_j$ is uniformly random among these three pages, the expected number of faults on $p^{\ell-1}_j$ during layer $\ell$ is at least $N/2$. Again, since $p^{\ell-1}_j$ is uniformly random among $\{p^{\ell}_{3j},p^{\ell}_{3j+1},p^{\ell}_{3j+2}\}$, by linearity of expectation we have,
  \[\mathbb{E}[X^{\ell-1}_j]\geq\frac{\mathbb{E}[X^{\ell}_{3j}]+\mathbb{E}[X^{\ell}_{3j+1}]+\mathbb{E}[X^{\ell}_{3j+2}]}{3}+\frac{N}{2}\text{.}\]
  By the induction hypothesis, each of $\mathbb{E}[X^{\ell}_{3j}]$, $\mathbb{E}[X^{\ell}_{3j+1}]$, $\mathbb{E}[X^{\ell}_{3j+2}]$ is at least $(m-\ell)\cdot N/2$. Thus, $\mathbb{E}[X^{\ell-1}_j]\geq(m-\ell+1)\cdot N/2$, as required.
\end{proof}

Having proven Lemma~\ref{lem_lb_adv} and Lemma~\ref{lem_lb_alg}, we are ready to prove the claimed lower bound.

\begin{proof}[Proof of Theorem~\ref{thm_lb}]
  By Lemma~\ref{lem_lb_adv}, the cost of the adversary's solution to the random instance generated by the adversarial strategy is $m+N$ with probability one. Note that at the end of the adversarial strategy, we are left with the singleton set $L_0$ containing the page $p^0_0$. The number of faults of the algorithm on page $p^0_0$ is a lower bound on the algorithm's cost with probability one. Thus, the algorithm's expected cost is at least the expectation of the number of algorithm's faults on $p^0_0$. By Lemma~\ref{lem_lb_alg}, this quantity is $\mathbb{E}[X^0_0]\geq mN/2$. Thus, the ratio of the algorithm's expected cost to the adversary's cost is at least $mN/(2\cdot(m+N))$, which approaches $m/2=(\log_3n)/2$ as $N\rightarrow\infty$. Thus, the competitive ratio of any randomized algorithm for min-max paging is at least $(\log_3n)/2=\Omega(\log n)$.
\end{proof}



\section{A Fractional Algorithm for General Paging}

We study a general class of convex objective functions for the paging problem to arrive at a competitive algorithm for min-max paging. Let $x \in \mathbb{R}^{T}$ be the vector consisting of the variables $x_{p,j}$ in order of appearance in $\sigma$.
The objective functions $f: \mathbb{R}^{T} \to \mathbb{R}$ which we consider in this section have the following properties: (1) $f(0) = 0$; (2) $f(x)$ is a monotonically increasing function in $x$; (3) $\nabla f(x)$ is monotonically increasing in each coordinate; and (4) $f(x)$ has \emph{$q$-bounded growth}, i.e. there exists a positive integer $q$ such that for all $x \in \R_{+}^T$, $\langle \nabla f(x), x \rangle \le q f(x)$. In particular, any polynomial function of $x$ of degree $q$ will fulfill these requirements.



We formulate the general paging problem as an online convex program. Given a convex function $f: \R_{+}^n \to \R$ and a matrix $A \in \R^{m \times n}$, a general (offline) convex programming problem is to minimize $f(x)$ subject to $A x \ge \ones$ and $x \ge \zeros$.

In online convex programming, the rows of the constraint matrix $A$ are revealed one by one, corresponding to the request sequence $\sigma$. Upon receiving the $t$th row $A_{t}$ of the constraint matrix, the task of the algorithm is to increase the variables $x$ until the constraint $A_{t} x \ge 1$ is fulfilled. The algorithm is never allowed to decrease any of the variables in $x$.

In the fractional convex program for paging, we denote by $p_t$ the page requested in round $t$. Furthermore, we let $r(p,t)$ indicate the number of requests to page $p$ up to and including round $t$, and let $t(p,j)$ be the round during which the page $p$ is requested for the $j$'th time. As each round corresponds uniquely to a pair $(p,j)$, we have $\sum_{p} r(p,t) = T$.
We let $B(t) = \{p \in P | r(p,t) \ge 1 \}$ be the set of distinct pages encountered up to, and including, round $t$. The variables $x_{p,j}$ can now take values in the interval $[0,1]$ and indicate the fraction of the page the algorithm has removed from the cache between the $j$'th and $j+1$'st times it was requested.
Using this notation, the convex program for general paging looks as follows:
\begin{equation}\label{primal}
  \begin{array}{lll}
    \text{minimize} & f(x) \\ 
    \text{subject to} & \sum_{p \in P \setminus \{p_t\}} x_{p, r(p,t)} \ge |B(t)| - k & \forall t \in [T] \\
                    & 0 \le x_{p,j} \le 1 & \forall p \in [n], j \in [r(p,T)] \\
  \end{array}
\end{equation}
By using a convex objective function, this formulation generalizes prior work on online paging, including weighted paging \cite{bansalbn_12}.
Crucially, the box constraint $0 \le x_{p,j} \le 1$ means that the online convex programming framework of \cite{azarOnlineAlgorithms2016} can not be used to solve this program.

At the beginning of round $t$, we are given a new variable $x_{p_t,r(p_t,t)}$, which is initialized to $0$ along with the constraint $\sum_{p \in B(t) \setminus \{p_t\}} x_{p, r(p,t)} \ge |B(t)| - k$.  This constraint ensures that after each round $t$, at least $|B(t)|-k$ fractional page mass has been ejected, or, equivalently, at most $k$ fractional page mass is inside the cache.  We observe that the variable $x_{p,j}$ will only appear in the constraints corresponding to rounds $t \in \{t(p,j)+1, t(p,j)+2, \dots t(p,j+1)-1 \}$, i.e.~the variable $x_{p,j}$ does not appear in round $t(p,j)$ when it is requested. This is because we are not allowed to increase $x_{p,j}$ during this round, as page $p$ is required to be fully inside the cache in round $t(p,j)$, in order to serve the request.

In order to define a dual for the convex program~\ref{primal}, we will need the following definition:

\begin{Definition}
Given a request sequence $\sigma$ of length $T$ consisting of pages from the set $P$, we can uniquely, up to relabeling of pages, define a constraint matrix $A
\in \{0,1\}^{T \times T}$ as
\[ 
  A_{t,(p,j)} =
  \begin{cases} 1 & \text{if $t \in [t(p,j)+1, t(p,j+1) - 1]$}\\ 0 &
\text{otherwise.}
  \end{cases}
\]
In round $t$, we can determine all non-zero entries, as they only depend on the variables encountered up to round $t$.  Additionally, we can implicitly set the columns corresponding to future variables to $0$. If we order both the columns and rows by order of appearance, then the constraint matrix will be lower triangular, see Figure~\ref{fig:matrix} in the appendix.
\end{Definition}

  \begin{Definition}
    Let $f: \mathbb{R}_{+}^{T} \to \mathbb{R}$ be a convex function. The fenchel dual $f^{*}: \mathbb{R}_{+}^{T} \to \mathbb{R}$ of $f$ is defined as $f^*(y) = \sup_{w \in \R_+^T} \left( \langle w, y \rangle - f(w) \right),$
    where $\langle w, y \rangle = \sum_{i=0}^T w_i \cdot y_i$ denotes the Euclidean scalar product.
  \end{Definition}

  We need the following property of the Fenchel dual in the analysis of our algorithm:

\begin{restatable}{property}{fenchelmonotone}
\label{o0} 
The Fenchel dual $f^*(y): \R_{+}^T \to \R$ of a convex function $f$ is monotonically increasing in $y$.
\end{restatable}

The dual will consist of two sets of $T$ variables each, denoted by $y_t$ and $z_{p,j}$, respectively.  We let $y$ be the vector consisting of the $y_t$ ordered increasingly in $t$ and $z$ being the vector consisting of the $z_{p,j}$ ordered the same way as $x$.

We will use the following \emph{conjugate dual} $D(y,z)$ for our primal-dual algorithm. 
For the convex primal~\eqref{primal}, the conjugate dual is:
\begin{equation*}
  \begin{array}{ll}
    \text{maximize} & D(y,z) = \sum_{t=1}^T (|B(t)| - k) y_t -\sum_{p, j} z_{p,j} - f^*(A^\top y - z), \\
    \text{subject to} &
                        \begin{array}{ll}0 \le y_t & \forall t \in [T]\\
                          0 \le z_{p,j} & \forall p \in [n], j \in [r(p,T)]\,,\\
                        \end{array}
  \end{array}
\end{equation*}

This dual differs from the dual used in \cite{bansalbn_12} by the inclusion of the Fenchel dual term $f^*(A^\top y - z)$ and from the dual used in \cite{azarOnlineAlgorithms2016} by the use of non-uniform coefficients for the $y_t$ variables and the inclusion of the variables $z_{p,j}$.
The dual will only be used to obtain a lower bound on $\OPT$ for the analysis of our algorithm, and it does not influence the primal solution the algorithm produces. It remains to show that the stated dual fulfills this property for the convex program~\eqref{primal}:

\begin{restatable}[Weak Duality]{lemma}{weakduality}
\label{weakduality}
  For any feasible $x \in [0,1]^T$ and $y, z \in \R_{+}^T$, we have
  \begin{equation*}
    f(x) \ge \sum_{t \in [T]} (|B(t)|-k) y_t - \sum_{p,j} z_{p,j} - f^*(A^\top y - z).
  \end{equation*}
\end{restatable}

Our online algorithm, given in Algorithm \ref{fractional} uses a continuous time $\tau$, which is 0 initially and increases throughout the algorithm. Let $\tau(t)$ denote the value of $\tau$ when we finish processing the $t$'th constraint and let $\tau(0) = 0$. As all variables are $0$ at creation and increase at a rate dependent on $\tau$, we use $x_{p,j}(\tau)$, $y_t(\tau)$, $z_{p,j}(\tau)$ to denote the values of the variables $x_{p,j}$, $y_t$, $z_{p,j}$ respectively at time $\tau$. Let $t(\tau)$ be the unique $t$ such that $\tau\in[\tau(t-1),\tau(t))$.
%
The algorithm uses parameters $r$ and $s_{p,j}(\tau)$ fixed later.  Note that $s_{p,j}(\tau)$ depends on the value of $\tau$, $p$ and $j$ and, thus, is not a constant parameter.

Our algorithm maintains dual variables $y$ and $z$ such that $A^\top y - z \le \delta \nabla f(x)$ is approximately fulfilled,~i.e.~, $A^\top y - z \le r \delta \nabla f(x)$ for some constant $r \ge 1$, which will then appear in the competitive ratio.  
We observe that if $f(x) = c^\top x$, the gradient is $c$ and we get $A^\top y - z \le c$, which is the dual constraint in the linear program for weighted paging. The reason why this point-wise upper bound is necessary is because, together with Property~\ref{o0}, it allows us to upper-bound the convex conjugate term $f^*(A^\top y - z)$ in $D(y,z)$ in terms of the primal function $f(x)$.  For general $w$ the conjugate $f^*(w)$ may be arbitrarily large as it is a convex function in $w$.

\begin{algorithm}
  \caption{A fractional algorithm for min-max paging}\label{fractional}
  \begin{algorithmic}[1]
    \Require $r > 0$ and $s_{p, j}(\tau) > 0$, $s_{p,j}(\tau)$ monotonically decreasing in $\tau$ 
    \State $\tau \gets 0$

    \For{each round $t \in [T]$}

    \State let $p_t$ be the page requested in this round 
    \State $x_{p_t,r(p,t)}(\tau) \gets 0$, $y_t(\tau) \gets 0$, $z_{p_t,r(p,t)}(\tau) \gets 0$


    \State{$\frac{d y_{t}(\tau)}{d\tau} \gets r$}

    \State{$\frac{d x_{p,r(p,t)}(\tau)}{d\tau} \gets \begin{cases} s_{p,r(p,t)}(\tau) \left(x_{p,r(p,t)}(\tau) + \frac{1}{k} \right) & \text{if } x_{{p,r(p,t)}}(\tau) < 1 \\ 0 & \text{otherwise} \end{cases}$}

    \State{$\frac{d z_{p,r(p,t)}(\tau)}{d\tau} \gets \begin{cases} r & \text{if } x_{{p,r(p,t)}}(\tau) = 1 \\ 0 & \text{otherwise} \end{cases}$}

    \State{Increase $\tau$, $y_{t}(\tau)$, $x_{p,r(p,t)}(\tau)$ and $z_{p,r(p,t)}(\tau)$ for all $p \in B(t) \setminus \{p_{t}\}$ simultaneously as per the above differential equations until $\sum_{p \in B(t) \setminus \{p_t\}} x_{p,r(p,t)}(\tau) \ge |B(t)| - k$}










    \EndFor
\end{algorithmic}
\end{algorithm}



Next, to bound the conjugate term in the dual, it is necessary to obtain a bound on the dual ``constraints" $A^\top y - z$, which we obtain by relating the constant growth of $y_t$ and $z_{p,j}$ to the exponential growth of the $x_{p,j}$:

\begin{restatable}{lemma}{xlowerbound}
  \label{xlowerbound} 
  Let $\bar{x}$ denote the value of $x$ after processing the complete request sequence $\sigma$, and similarly for $\bar{y}$ and $\bar{z}$.  If $s_{p,j}(\tau)$ is monotonically decreasing in $\tau$, then
  \begin{equation}
    \label{lem2eq}
    \bar{x}_{p,j} \ge \frac1k \left(\exp \left(
        \frac{ s_{p,j}'}{r} \left( \sum_{t = t(p, j)+1}^{t(p,j+1)-1} \bar{y}_t -
          \bar{z}_{p,j} \right) \right) - 1\right),
  \end{equation}
  where $s_{p,j}'$ is the minimum value that $s_{p,j}(\tau)$ takes on during the execution of the algorithm.
\end{restatable}

The following is an immediate consequence of the previous lemma and the fact that $x_{p,j} \le 1$:

\begin{restatable}{lemma}{feasibility}
    
\label{feasibility1}
The $x(\tau(t))$ produced by Algorithm \ref{fractional} throughout its execution are feasible for the primal for all $t \in [T]$ and the vector $(A^\top \bar{y} - \bar{z})_{p,j} \le \frac{r}{s_{p,j}'} \ln(k+1)$.
\end{restatable}

%
%

The conjugate of a convex function with bounded growth can be bounded in terms of the original function and $q$, using the following lemma:

\begin{lemma}[\cite{azarOnlineAlgorithms2016}]\label{conjugateabc}
Let $f: \mathbb{R}_{\ge 0}^T \to \mathbb{R}_{\ge 0}$ be a monotone, convex, differentiable function satisfying $f(0) = 0$.  If there is a $q > 1$ such that $\langle \nabla f(x), x \rangle \le q f(x)$, then for any $0 < \gamma < 1$, $y \in \mathbb{R}_{\ge 0}^T$, $f^*(\gamma y) \le \gamma^{\frac{q}{q-1}} \cdot f^*(y)$ and $f^*(\gamma \nabla f(y)) \le \gamma^{\frac{q}{q-1}} (q-1) f(y)$.
\end{lemma}

\begin{theorem}\label{thm:frac}
  Let $f(x)$ be a convex function satisfying the requirements stated at the beginning of this section, and let $\sigma$ be any request sequence. If we set $s_{p,j}(\tau) = \frac{\partial f(x)}{\partial x_{p,j}}^{-1}$ and $r = \frac{1}{\ln(k+1)(2q\ln(k+1))^{q-1}}$, then Algorithm~\ref{fractional} produces a $(2q \log(k+1))^q$-competitive solution $\bar{x}$ for fractional paging with objective function $f(x)$ in an online manner.
    
\end{theorem}
\begin{proof}
  By weak duality, it suffices to show that the primal is no larger than $O((q \log(k+1))^q)$ times the dual, which is a lower bound on the cost of an optimal solution $x^*$ by weak duality.
  We will bound the primal and the dual growth rates for each round $t$. It suffices to only consider the case $A_t x(\tau) < |B(t)| - k$, as otherwise, the round is finished, and nothing needs to be done. 
  
  The processing of round $t$ begins at time $\tau(t - 1)$ and will last until $\tau(t)$, so we assume that $\tau \in (\tau(t-1), \tau(t)]$ for the remainder of this proof.
%
  Let $C_t(\tau) = \{ (p,j) \mid t(p,j) < t < t(p,j+1) \text{ and } x_{p,j}(\tau) < 1 \}$ be the set of indices of variables $x_{p,j}(\tau)$ in round $t$ which correspond to a page that is (partially) in the cache, i.e., the indices of the variables corresponding to the latest request of a given $p$, which are increasing and have not been fully removed from the cache.
  Similarly, let $D_t(\tau) = \{ (p,j) \mid t(p,j) < t < t(p,j+1) \text{ and } x_{p,j}(\tau) = 1 \}$ be the set of indices of the variables $x_{p,j}(\tau)$ which have been fully removed from the cache since they have been last requested and which correspond to the latest request to a given page $p$.  Note that $|C_t(\tau)| + |D_t(\tau)| = |B(t)| - 1$, as the sets $C_t(\tau)$ and $D_t(\tau)$ are disjoint and include a variable for each page except the page $p_t$.
While processing the $t$-th constraint, we have, by the choice of $s_{p,j}(\tau)$, and the fact that $x_{p,j}(\tau)$ is constant if $(p,j) \in D_t(\tau)$:
  \begin{align}\label{g1}
    G_1 \coloneqq \frac{df(x)}{d\tau} = \sum_{p,j} \frac{\partial f(x)}{\partial x_{p,j}} \frac{\partial x_{p,j}(\tau)}{\partial \tau}
    &=\sum_{(p,j) \in C_t{(\tau)}} \frac{\partial f(x)}{\partial x_{p,j}} \left( s_{p,j}(\tau) \left(x_{p,j}(\tau) + \frac1k \right) \right) \\
    &= \sum_{(p,j) \in C_t{(\tau)}} \left(x_{p,j}(\tau) + \frac1k \right) \le |B(t)| - k - |D_t(\tau)| + \frac{|C_t(\tau)|}{k}.\label{eq7}
  \end{align}
  The first equality is due to the chain rule for vector-valued functions. The second equality uses the definition of $\frac{\partial x_{p,j}(\tau)}{\partial \tau}$ and the fact that $x_{p,j}$ does not change for $(p,j) \notin C_t{(\tau)}$.  And, the last inequality follows from the fact that the variables in $D_t(\tau)$ are all equal to $1$.

Note that only the $y_t$ corresponding to round $t$ may increase during round $t$. For the linear term $\sum_{t} (|B(t)|-k)y_t(\tau) - \sum_{p,j} z_{p,j}(\tau)$ in the dual, it holds that in round $t$
  \begin{align}\label{g2} 
  G_2 \coloneqq \frac{d}{d\tau} \left( \sum_{t} (|B(t)|-k)y_t(\tau) - \sum_{p,j}
    z_{p,j}(\tau) \right) = (|B(t)| - k) r - \sum_{(p,j) \in D_t(\tau)} r = r
    (|B(t)| - k - |D_t(\tau)|).
  \end{align}
  
We note that the right-hand side of Equation~\eqref{g2} is $r$-times the first term of the right-hand side of Equation~\eqref{eq7}.
%
Furthermore, 
$\frac{|C_t(\tau)|}{k} \le |C_t(\tau)| - k + 1 = |B(t)| - 1 - |D_t(\tau)| - k + 1 = \frac1r G_2,$ since $|C_t(\tau)| \ge k$. 
%
By adding together Equation~\eqref{g2} and the last inequality, we obtain
\begin{equation}\label{eq11}
G_1 \le |B(t)| - k  - |D_t(\tau)| + |C_t(\tau)|/k \le G_2/r  + G_2/r = 2 G_2/r.
\end{equation}


Since both the primal and the linear term of the dual initially have value $0$ at time $\tau = 0$, their overall competitive ratio after processing all elements will be $\frac2r$. 
  Thus for the choice of $r(\delta) = \frac{\delta}{\ln(k+1)}$, where $\delta$ is a parameter which we will optimize later,
  from Equation~\eqref{eq11} it follows that $\sum_t (B(t)-k) \bar{y}_t - \sum_{p,j} \bar{z}_{p,j} \ge \frac{\delta}{2\ln(k+1)} f(\bar x).$
  Plugging $s_{p,j}(\tau) = \frac{\partial f(x)}{\partial x_{p,j}}^{-1}$ and
  $r(\delta) = \frac{\delta}{\ln(k+1)}$ into the second statement of Claim \ref{feasibility1}, we obtain that $A^T y - z \le \delta \nabla f(\bar{x})$, which allows us to bound the conjugate term of the dual as
  \[
    f^*(A^T \bar y - \bar z) \le f^*(\delta \nabla f(\bar x)) \le 
    \begin{cases}
      \delta^{\frac{q}{q-1}} \cdot (q - 1) \cdot f(\bar x) & \text{ if $q > 1$},\\
      0 & \text{ if $q = 1$},
    \end{cases}
  \]
 where the first inequality is due to Property~\ref{o0} and the second inequality uses Lemma~\ref{conjugateabc}.  Hence the relationship between the final value $D(\bar y, \bar z)$ of the dual and the final value $f(\bar x)$ of the primal is

  
  \[
    D(\bar y, \bar z) = \sum_t (|B(t)|-k) \bar{y}_t - \sum_{p,j} \bar{z}_{p,j} -
    f^*(A^T \bar y - \bar z) \ge \left( \frac{\delta}{2 \ln(1+k)} -
      \delta^{\frac{q}{q-1}} \cdot (q - 1)\right) \cdot f(\bar x).
  \]
  The term $h(\delta) = \left( \frac{\delta}{2 \ln(k+1)} - \delta^{\frac{q}{q-1}} \cdot (q - 1)\right)$ is a polynomial in $\delta$, which governs our competitive ratio. The best competitive ratio is obtained if we find $\delta \in (0,1)$ such that $h(\delta)$ is maximized. We find a local maximum at $\delta^* = \frac{1}{(2 q \ln(k + 1) )^{q-1}}$, yielding $h(\delta^*) = \frac{1}{(2 q\ln(k+1))^q}$.  By rearranging and weak duality (Lemma~\ref{weakduality}) we obtain
  \[
    f(\bar{x}) \le \left(2 q \ln(k+1) \right)^q D(\bar y, \bar z) \le \left(2 q \ln(k+1) \right)^q f(x^*),
  \]
  where $x^*$ is an optimal solution.
\end{proof}




The $\ell_q$-norm does not lie in our class of objective functions, as a coordinate of $\nabla f(x)$ can decrease while we increase all coordinates of $x$, hence we can not apply Theorem~\ref{thm:frac} straight away.

\begin{theorem}\label{qpaging}
Let $q \in [1, \infty)$.  Then there exists a $2q \log(k+1)$-competitive algorithm for fractional $q$-paging with a cache of size $k$.
\end{theorem}
\begin{proof}
Let us fix $q \in [1, \infty)$. We apply Theorem~\ref{thm:frac} with the target function $f(x) = \sum_{p \in P} \left(\sum_{j=1}^{r(p,T)} x_{p,j}\right)^q$, which is the $q$th power of the $\ell_q$-norm. This produces a solution $\bar{x}$, which is $(2q \log(k+1))^q$-competitive for the paging problem with target function $f(x)$.

Let $g: \mathbb{R} \to \mathbb{R}$ be a monotone function, then a solution $x$ to the paging problem with target function $f(x)$ will also be a feasible solution to the paging problem with target function $g(f(x))$. In particular, as $g$ preserves the standard ordering on the reals, an optimal solution to paging with target function $f(x)$ will remain an optimal solution to the problem with target function $g(f(x))$.

If we let $g(y) = y^{\frac{1}{q}}$ and we let $x^*$ be an optimal solution to the paging problem with target function $f(x)$, then $g(f(x))$ will be the $\ell_q$-norm and we find that $g(f(\bar{x})) \le g((2q \log(k+1))^q f(x^*)) = 2q \log(k+1) g(f(x^*))$.
\end{proof}

\begin{Remark} 
Note that if the gradient of $f(x)$ is $0$ at $x=0$, then we start the algorithm at $\epsilon \cdot \mathbf{1}$ for a small $\epsilon > 0$ instead, which can be chosen sufficiently small, so it does not influence the competitive ratio.
\end{Remark}




We use
Theorem~\ref{qpaging} to show that we can obtain a $2e\log(n) \log(k+1)$-competitive fractional solution for $\infty$-paging by reducing it to $\log n$-paging.

\begin{theorem}\label{thm_frac_ub}
  There exists a $2e \log(n) \log(k+1)$-competitive algorithm for fractional
  min-max paging.
\end{theorem}



\begin{proof}
Let $x^*$ be the optimal solution to the $\infty$-paging problem for the request sequence $\sigma$. We denote the cost of this solution by $\OPT_\infty$. 
%
Let $\bar{x}$ denote the fractional solution obtained using Algorithm~\ref{fractional}. By Theorem~\ref{qpaging} and $\|x\|_\infty \le \|x\|_{\log n} \le e \|x\|_\infty$, we know that this solution has cost
\[
\max_{p \in P} \sum_{j=1}^{r(p,T)} \bar{x}_{p,j} \le \left(\sum_{p \in P} \left(\sum_{j=1}^{r(p,T)} \bar{x}_{p,j} \right)^{\log n} \right)^{\frac{1}{\log n}} \le 2 \log(n) \log(k+1) \cdot \OPT_{\log n} \le 2e \log(n) \log(k+1) \cdot \OPT_{\infty},
\]
where $\OPT_{\log n}$ denotes the cost of an optimal solution to the $\log n$-paging problem with input $\sigma$.
%
%
This implies  that $\bar{x}$ is a $2e \log(n) \log(k+1)$-competitive solution for $\infty$-paging.
\end{proof}

\section{Rounding Fractional Solutions Online}\label{sec:round}

\subsection{\texorpdfstring{An $O(k \log(n) \log(k))$-competitive Deterministic Algorithm}{An O(k log(n) log(k))-competitive Deterministic Algorithm}}

This section shows how to round a fractional solution for min-max paging to an integral solution online. The rounding procedure is deterministic and, when coupled with a fractional min-max paging algorithm, gives a deterministic min-max paging algorithm.

\begin{theorem}\label{thm_rounding}
If there exists an $\alpha$-competitive algorithm for fractional min-max paging with cache size $k$, then there exists a $(\alpha k)$-competitive deterministic algorithm for min-max paging with cache size $k$.
\end{theorem}

\begin{proof}
Without loss of generality, we assume that the fractional min-max paging algorithm is lazy. That is, it loads a page only when the page is requested. Indeed, an arbitrary solution can be converted into a lazy solution online without increasing the cost by delaying page loads as much as possible.

The deterministic integral algorithm maintains the following invariant: it always has a page $p$ in its cache whenever the fractional algorithm has more than a $1-1/k$ fraction of $p$ in its cache. We observe that the fractional algorithm must always fully have at least one page in its cache: the most recently requested page. Therefore, at any time, the number of pages $p$ such that the fractional algorithm contains more than a $1-1/k$ fraction of $p$ is less than $1+(k-1)/(1-1/k)=k+1$, and therefore, this number is at most $k$.

Consider an arbitrary request to some page $p$. If the integral algorithm already has $p$ in its cache, it ignores the request, whereas the fractional algorithm possibly serves the request by evicting some pages fractionally. On the other hand, suppose the integral algorithm does not already have $p$ in its cache, then this implies that the fractional algorithm has at most a $1-1/k$ fraction of $p$ in its cache. After the fractional algorithm brings $p$ into its cache, the integral algorithm must have a page $q$ in its cache such that the fractional algorithm has at most a $1-1/k$ fraction of $q$ in its cache. (Otherwise, the fractional algorithm has more than a $1-1/k$ fraction of $k+1$ pages in its cache, namely, the $k$ pages in the integral algorithm's cache and the page $p$, thus contradicting the observation from the last paragraph.) The integral algorithm replaces one such page $q$ by $p$ to serve the request and thus, maintains the invariant. In this process, the integral and the fractional algorithms incur $1$ and at least $1/k$ faults, respectively, on page $p$.

Thus, at the end of the request sequence, for every page $p$, the number of faults of the integral algorithm on $p$ is at most $k$ times the number of faults of the fractional algorithm on $p$. Thus, the cost of the integral algorithm is at most $k$ times the cost of the fractional algorithm. Since the latter is at most $\alpha$ times the cost of the optimum, the cost of the integral algorithm is at most $\alpha k$ times the cost of the optimum solution.
\end{proof}

\begin{corollary}\label{detround}
There exists a $2ek\log(n)\log(k+1)$-competitive deterministic algorithm for min-max paging.
\end{corollary}

\begin{proof}
Follows from Theorem~\ref{thm_frac_ub} and Theorem~\ref{thm_rounding}.
\end{proof}

It is noteworthy that the trick in the proof of Theorem~\ref{thm_rounding} can also be used for the derandomization of randomized algorithms. Specifically, suppose an $\alpha$-competitive randomized algorithm exists for min-max paging. Then there also exists a fractional one with the same competitive ratio. Thus, by Theorem~\ref{thm_rounding}, there exists a $\alpha k$-competitive deterministic algorithm for min-max paging.

\subsection{\texorpdfstring{An $O(\log^{2}(n) \log(k))$-competitive Randomized Algorithm.}{An O(log2(n) log(k))-competitive Randomized Algorithm.}}

Using a more sophisticated rounding approach, we obtain a randomized algorithm whose competitive ratio no longer depends linearly on $k$, in exchange for an additional $\log(n)$ factor. This result rules out a lower bound of $\Omega(k)$. This algorithm is of interest in the regime where $\log(n) \le k$, which is often the case in applications.


We can obtain a randomized algorithm for min-max paging by using the rounding scheme for weighted paging of Bansal et al.~\cite{bansal2021efficient}.
The simplified rounding scheme is presented in Algorithm~\ref{randround}. 
Each online rounding step only depends on the previous, and current fractional cache states $x(t-1)$ and $x(t)$
as well as the previous integral cache state and  on a parameter $\beta$, which indicates how aggressively we eject pages from the cache.
\emph{The rounding scheme works for any caching scheme that fulfills the condition that (1) at any time $t$, for any page $p \ne p_r$, $x_p(t) - x_p(t-1) \ge 0$ and (2) the total fraction of pages evicted upon any request is at most 1.
} Algorithm~\ref{fractional} indeed has these properties, so we can use the rounding scheme as long as we can relate the rounding costs to our target function, even though we solve a different paging problem than they do.

Let $x$ be a fractional solution produced by Algorithm~\ref{fractional}. After processing round $t$, the algorithm will produce a fractional value $x_p(t)$ for each page, indicating the fraction of page $p$ in the cache in this round. In other words, the process of solving the fractional problem online produces, whenever Algorithm~\ref{fractional} finishes processing a round at time $\tau(t)$, the vector
\[
  x(t) = \begin{bmatrix} x_{p_{1},r(p,t)}(\tau(t)) \\ \vdots \\ x_{p_{n},r(p,t)}(\tau(t)) \end{bmatrix}.
\]

We let $y_p(t) = \min\{\beta \cdot x_p(t), 1\}$ be the solution in which every coordinate is scaled up by a factor of
$\beta$. The factor $\beta$ governs how much more aggressively pages are evicted from the cache.

Algorithm~\ref{randround} may evict pages and incur costs in two separate places. The first type we need to account for is the cost incurred via the random evictions of pages in the for-loop in lines 4-5 of the algorithm. The second type is the cost incurred by fixing the cache size in lines 6-7 if no page was evicted in the for-loop. We will bound these costs separately and combine them in our upper bound.

\begin{algorithm}
  \caption{The randomized rounding scheme of~\cite{bansal2021efficient} adapted to our problem.}\label{randround}
  \begin{algorithmic}[1]
    \Procedure{Round}{$x(t), x(t-1), C(t - 1)$}
    \If{$p_{t} \notin C(t - 1)$}\Comment{Add the page $p_{t}$ to the cache, if it is not already in it.}
    \State $C(t - 1) \gets C(t-1) \cup \{p_{t}\}$
    \EndIf

    \For{$p \in C(t-1) \setminus \{p_{t}\}$}
    \State Evict $p$ from $C(t-1)$ independently with probability $\frac{y_{p}(t) - y_{p}(t-1)}{1 - y_p(t-1)}$
    \EndFor

    \If{$|C(t-1)| > k$}
    \State Evict an arbitrary page $p \neq p_{t}$ from $C(t-1)$
    \EndIf

    \State $C(t) \gets C(t-1)$
    \EndProcedure
  \end{algorithmic}
\end{algorithm}

For the first type, it is easy to see that the cost incurred for evicting a page in lines 4-5 depends only on the sequence of fractional values $y_{p}(1), y_{p}(2), \dots, y_{p}(T)$ that this page takes on and it is independent of the values $y_{p'}(t)$ for all $t$ and $p'\neq p$. In particular, the probability $y_{p,j}$ that a page is evicted
in lines 4-5, between its $j$-th and $j+1$-st request is
\begin{align*}
    \sum_{t=t(p,j)+1}^{t(p,j+1)-1} \Pr[\text{page $p$ is evicted in round $t$}] &=
    \sum_{t=t(p,j)+1}^{t(p,j+1)-1} \frac{y_p(t) - y_p(t-1)}{1 - y_p(t-1)} \Pr[\text{page $p$ is not evicted until round $t$}]\\
    & = \sum_{t=t(p,j)+1}^{t(p,j+1)-1} \frac{y_p(t) - y_p(t-1)}{1 - y_p(t-1)} \prod_{t'=t(p,j)+1}^{t-1} 1 - \frac{y_p(t') - y_p(t'-1)}{1 - y_p(t'-1)}\\
    & = \sum_{t=t(p,j)+1}^{t(p,j+1)-1} \frac{y_p(t) - y_p(t-1)}{1 - y_p(t-1)} \prod_{t'=t(p,j)+1}^{t-1} \frac{1 - y_p(t')}{1 - y_p(t'-1)} \\
    & = \sum_{t=t(p,j)+1}^{t(p,j+1)-1} \frac{y_p(t) - y_p(t-1)}{1-y_p(t(p,j))} = \frac{y_p(t(p,j+1)-1)}{1-y_p(t(p,j))} = y_{p}(t(p,j+1)-1).
\end{align*}
The second equation holds because of the independence of the probability of eviction in different rounds; the fourth holds because it is a telescoping product, and the last equation holds as $y_p(t(p,j)) = 0$.
Let $Y_{p,j}$ be a Bernoulli random variable that is $1$ with probability $y_p(t(p,j+1)-1)$ and let $Y_{p} = \sum_{j} Y_{p,j}$ be the sum of all $Y_{p,j}$ for fixed $p$. We let these variables track the expected cost of evictions for each page. By linearity of expectation, we immediately see that
\[
\mathbb{E}[Y_p] = \mathbb{E} \left[ \sum_{j=1}^{r(p,T)} Y_{p,j} \right] = \sum_{j=1}^{r(p,T)} \Pr[\text{$p$ is evicted between request $j$ and $j+1$}] = \sum_{j=1}^{r(p,T)} y_{p,j} \le \beta  \sum_{j=1}^{r(p,T)} x_{p,j}.
\]
It follows that
\[
\max_p \mathbb{E}[Y_p] \le \beta \max_p \sum_{j=1}^{r(p,T)} x_{p,j},
\]
where the right-hand side is $\beta$ times the cost of the fractional solution $x$.
It remains to relate the left side of this inequality with $ \mathbb{E}[\max_{p} Y_{p}]$.

\begin{lemma}
  Let $Y_{p,j}$ be Bernoulli random variables which are $1$ with probability $y_{p,j}$. Let $Y_p = \sum_{j} Y_{p,j}$ and assume there are $n$ such sums, then
  \[
    \mathbb{E}[\max_{p} Y_{p}] \le e \cdot \max_{p} E[Y_{p}] + \log(n).
  \]
\end{lemma}
\begin{proof}
  Let $Y = \max_{p} Y_{p}$. Using Jensen's inequality, we get the first inequality in the following chain of inequalities:
  \begin{align*}
    \exp(\mathbb{E}[Y]) &\le \mathbb{E}[\exp(Y)] = \mathbb{E}[\max_{p} \exp(Y_{p})] \\
    & \le \sum_{p} \mathbb{E}[\exp(Y_{p})]
    = \sum_{p} \prod_{j=1}^{r(p,T)} \mathbb{E}[\exp(Y_{p,j})] \\
    &= \sum_{p} \prod_{j=1}^{r(p,T)} (1 - y_{p,j} + e y_{p,j})
    \le \sum_{p} \prod_{j=1}^{r(p,T)} e^{e \cdot y_{p,j}}\\
    &= \sum_{p} e^{e \cdot \sum_{j=1}^{r(p,T)} y_{p,j}} \le n \cdot \max_{p} e^{e \cdot \mathbb{E}[Y_{p}]}.
  \end{align*}
  The first equality follows as $\exp$ is a monotone function, and the second equality follows by the independence of the $Y_{p,j}$.
  After taking logarithms, we obtain
  \[
    \mathbb{E}[Y] \le e \cdot \max_{p} \mathbb{E}[Y_{p}] + \log(n),
  \]
  which yields the desired result.
\end{proof}

Therefore, by the above lemma, the expected cost of the first type of costs is bounded by
\begin{equation}\label{type1costs}
 \mathbb{E}[\max_{p} Y_{p}] \le  O(1) \cdot \beta \max_p \sum_{j=1}^{r(p,T)} x_{p,j}.
\end{equation}

\begin{lemma}[\cite{bansal2021efficient}]\label{evictions}
Let $x$ be a fractional solution for a general paging problem. The expected cost of resets is at most $16 k e^{-\beta/4} \cdot\sum_{p \in P} \sum_{j=1}^{r(p,T)} x_{p,j}$.
\end{lemma}

By choosing $\beta = 4\log(nk)$ in Lemma~\ref{evictions}, the expected total cost of resets for the solution $y$ becomes
\begin{equation}\label{type2costs}
16k e^{-\log(nk)} \sum_{p \in P} \sum_{j=1}^{r(p,T)} y_{p,j} \le \frac{16}{n} \sum_{p \in P} \sum_{j=1}^{r(p,T)} y_{p,j} \le \frac{16\beta}{n} \sum_{p\in P} \sum_{j=1}^{r(p,T)} x_{p,j} \le 16 \beta \max_{p \in P} \sum_{j=1}^{r(p,T)} x_{p,j}
\end{equation}
where the last inequality follows due to the fact that the average cost per page $\frac1n \sum_{p\in P} \sum_{j=1}^{r(p,T)} x_{p,j}$ is a lower bound on the maximum cost of a single page in the solution $x$.

Crucially, Lemma~\ref{evictions} depends on the following helper lemma: 

\begin{restatable}{lemma}{roundinghelper}
\label{roundinghelper}
  Given a fractional solution $x$ to the paging problem, we can find a fractional solution $x^{*}$ in which every variable is a multiple of $\delta = \frac{1}{4k}$, and the cost of which is no more than $3$ times the cost of $x$.
\end{restatable}


Taking the bounds on the two types of costs, namely Equations~\eqref{type1costs}~and~\eqref{type2costs}, we have shown the following:

\randomizedfinal*



\section{Further related work}\label{sec:related-work}

Sleator and Tarjan~\cite{SleatorT85} defined the framework of online algorithms and competitive analysis, and paging is one of the earliest problems studied in the online setting. Several deterministic algorithms, such as ``Least Recently Used'' (LRU) and ``First In First Out'' (FIFO), among others, are known to achieve the optimal deterministic competitive ratio of $k$~\cite{SleatorT85}, where $k$ is the maximum number of pages that can be inside the cache at any point in time. The randomized competitive ratio is known to be $H_k$, where the upper bound is due to Achlioptas et al.~\cite{AchlioptasCN_TCS00} and the lower bound is due to Fiat et al.~\cite{FiatKLMSY_JAlg91}.

Several practical generalizations of the paging problem have been studied and they are known to have a deterministic competitive ratio of $k$~\cite{ChrobakKPV90,Young98} and randomized competitive ratio $\Theta(\log k)$~\cite{bansalbn_12,BansalBN12_sicomp}. These include weighted paging -- where pages have arbitrary loading costs, the \textit{bit model} -- where pages have arbitrary sizes and loading cost proportional to size, the \textit{fault model} -- where pages have arbitrary sizes but unit loading cost, and generalized paging -- where pages have arbitrary loading costs as well as sizes. Interestingly, all these results are robust in the sense that they all extend to the resource-augmentation setting, where the adversary has fewer servers than the algorithm. It is noteworthy that the line of work in search of a randomized algorithm for these paging variants by Bansal, Buchbinder, and Naor led to the development of the online primal-dual framework for designing fractional algorithms for online problems, whose solutions can often be rounded to an integral solution online.

A simple-looking but intriguing generalization of paging is the $k$-server problem defined by Manasse, McGeogh, and Sleator~\cite{ManasseMS88}, which concerns moving $k$ mobile servers on a metric space to serve requests while minimizing total movement. (The paging problem is the $k$-server problem on the uniform metric over the set of pages.) While Manasse et al.~\cite{ManasseMS88} proved a lower bound of $k$ on the deterministic competitive ratio for every metric space with more than $k$ points, the existence of a $k$-competitive algorithm is still unknown, and this is popularly called the $k$-server conjecture. The best-known $k$-server algorithm that works for all metrics called the \textit{Work Function Algorithm} by Koutsoupias and Papadimitriou~\cite{KoutsoupiasP95}, achieves a competitive ratio of $2k-1$. For randomized algorithms, surprisingly, neither a better upper bound than the deterministic $2k-1$ nor a better lower bound of $\Omega(\log k)$ arising from paging is known. Koutsoupias~\cite{Koutsoupias09} presents a more comprehensive discussion on the $k$-server problem.

\bibliographystyle{alpha}
\bibliography{ref}

\appendix

\section{Further Details for Lower Bounds}
\label{app:lb}
This section shows the complete details for the deterministic lower bound for min-max paging. We begin with a proof of the lower bound for $k=2$, which neatly highlights the core construction lying at the heart of the lower bound.


\begin{lemma}
  Any deterministic algorithm $\ALG$ for min-max paging with cache size $k = 2$ is at least $\Omega(\log n)$-competitive.
\end{lemma}
\begin{proof}
  Suppose $n = 3^\ell$ and let the set of pages be $\{p_{1}, p_{2}, \dots, p_{n}\}$. We construct a bad request sequence $\sigma$ for $\ALG$ in $\ell$ layers. Each layer is further divided into phases. Let $N \gg n$ be a large integer parameter.

  We call the number of page faults incurred by page $p$ up to round $t$ the \emph{cost} of $p$ at $t$. Similarly, the \emph{cost of the min-max paging algorithm $\ALG$} is the maximum cost over all pages $p \in P$.


  We now iteratively construct the adversarial sequence $\sigma$, going layer by layer.

  Layer~1 will use all pages, that is the set $\{p_{1},p_{2}, \dots, p_{n}\}$.
\begin{itemize}
  \item In the first phase, we request pages $p_{1}$, $p_{2}$, and $p_{3}$ in such a way that $\ALG$ faults on every request, such a cruel sequence composed of $k+1$ pages exists for every deterministic algorithm for paging. We stop this phase once the cost of the algorithm becomes $N$, which must happen before sending $3N$ requests. Without loss of generality, we assume that the cost of $p_{1}$ first reaches $N$, and hence the costs of $p_{2}$ and $p_{3}$ are $< N$. These costs are the same as the number of requests to the respective pages because the algorithm always faults.

  \item In the second phase, we repeat this step with pages $p_{4}$, $p_{5}$, and $p_{6}$.
  Without loss of generality, we assume that the cost of $p_{4}$ is $N$.
  \item Repeat this process until the set of pages $\{p_{1}, p_{2}, \dots, p_{n}\}$ is exhausted.
\end{itemize}
In all phases, the adversary always keeps the lowest numbered page, that is pages $p_{1}, p_{4}, p_{7}, \ldots$, respectively, in the cache, incurring a cost of only $1$ on them, while the cost of the other pages is at most $N$.  We \emph{promote} pages $p_{1}, p_{4}, p_{7}, \ldots$ to Layer~2.

Layer~2 with universe of pages $\{p_{1}, p_{4}, p_{7}, ..., p_{n-2}\}$ is constructed exactly in the same way as Layer~1. After this layer, the cost of $\ALG$ is $2N$, whereas the cost of the adversary is $\le N$, with costs of pages $p_{1}, p_{10}, p_{19},\ldots$ having cost $\le 2$, and we \emph{promote} them to Layer~3, and so on.

After phase $i$, the number of pages in the universe becomes $n/3^i$, the cost of $\ALG$ becomes $iN$, whereas the adversary's cost is always $\le \max\{N, i\}$.  This gives us the desired lower bound using $\log_3(n)$ layers by choosing $N \ge \ell$.
\end{proof}

\paragraph{Generalizing the above construction} The idea behind the lower bound of $\Omega(k (\log n)/\log k)$ is as follows.  We generalize the construction above by using $n = (k+1)^\ell$ pages and $\ell = \log_{k+1}n$ layers. In each phase, we use $k{+}1$ pages and force the cost of the algorithm on one of these pages to increase by $N$.  The adversary's cost increases by at most $1$ on the page she will promote to the next layer. By using a smarter \emph{offline} algorithm, the cost of the adversary increases by at most $O(N/k)$ on the $k$ pages of this phase that will not be promoted. So, in the end, the adversary's cost is $O(\ell + N/k)$, whereas the cost of the algorithm is $\Omega(N\ell)$. We obtain the desired lower bound by choosing $N \ge c k \ell$ for a large enough constant $c$.



For any paging algorithm ALG and any request sequence $\sigma$, we define $\cost(\ALG, \sigma, p, t)$ to be the the number of page faults incurred on page $p$ after processing the first $t$ requests of $\sigma$. Furthermore, we define \[\cost(\ALG, \sigma) = \max_{p \in P} \cost(\ALG, \sigma, p, T),\] to be the overall cost incurred by ALG while processing request sequence $\sigma$.

The optimal offline algorithm OPT for min-max paging is not known to us, so we use Algorithm~\ref{greedylfd} (GreedyLFD) to obtain an upper bound on the cost of OPT. Intuitively, this algorithm avoids increasing its maximum cost for as long as possible by greedily keeping the most expensive pages in its cache. As GreedyLFD is an offline algorithm, it has access to the complete request sequence $\sigma$ and can always eject the page that is \emph{furthest in the future}. That is, in round $t$, it ejects the page $p \in C$ whose next occurrence comes last in the remainder of $\sigma$ after $t$. If a page does not occur in the remainder of $\sigma$, it is treated as being infinitely far in the future, and the algorithm will always prefer to eject this page over one that will still occur in $\sigma$.

\begin{algorithm}
  \caption{The offline algorithm GreedyLFD}\label{greedylfd}
  \begin{algorithmic}[1]
    \Procedure{GreedyLFD}{$\sigma$} \State $C \gets \emptyset$
    \Comment{Initialize the cache.}  \State $t \gets 1$ \For{$p \in P$} \State
    $c_p \gets 0$ \Comment{Counter variables for the number of faults on page
      $p$. At any point in time $c_p = cost(GreedyLFD, \sigma, p, t)$.}  \EndFor
    \While{$t < T$} \If{the $t$th element, say $p_t$, of $\sigma$ is not in $C$}
    \State $c_p \gets c_p + 1$ \If{$|C| < k$} \State $C \gets C \cup \{p\}$
    \Else \State $S \gets \{q \in C \mid c_q < \max_r c_r \}$ \Comment{Obtain
      the set of pages whose cost is less than the current maximum.}

    \If{$S = \emptyset$} \Comment{This happens if all pages in $C$ are of the
      same cost.}  \State $S \gets C$ \EndIf

    \State Evict $q \in S$ which next occurs farthest in the future.  \EndIf
    \EndIf \State $t \gets t+1$ \EndWhile \EndProcedure
  \end{algorithmic}
\end{algorithm}

\begin{lemma}\label{lfdlb}
  Let $\sigma$ be a request sequence for min-max paging using $k+1$ unique  pages. Then
  \[
    \cost(GreedyLFD, \sigma) \le \frac{2 (\len(\sigma) - 2k - 1)}{2k + k
      (k+1)}+2.
  \]
\end{lemma}
\begin{proof}
  We fix $\sigma$ to be an arbitrary request sequence of length $T$ using pages from the set $P = \{p_1, p_2, \dots, p_{k+1}\}$. Let
  \[
  t_i = \min \left\{t \in [T] \mid \max_{p \in P} \cost(GreedyLFD, \sigma, p,t) \ge i \right\}
  \]
  be the first time GreedyLFD faults on a page for the $i$th time. We note that at time $t_i$, there is only one page of cost $i$.

  Furthermore, we note that $t_1 = 1$, as the algorithm starts with an empty cache and $t_2 \ge 2k+1$, as the algorithm will fault on the first $k+1$ distinct pages it encounters, and it will then eject the page of cost $1$ which occurs farthest in the future. As there are $k$ pages in its cache, at least one page will not occur in the next $k-1$ time steps, so the shortest sequence that can cause GreedyLFD to fault two times on a single page is of length $2k+1$.

  We now show $t_{i+1} \ge t_i + k + \frac{k(k+1)}{2}$ for all $i \ge 2$. Let us fix $i \geq 2$ and assume we are currently at time $t_{i}$. This means that there exists some page $p \in P$ for which $\cost(GreedyLFD, \sigma, p, t_i) = i$ and it is the only page of cost $i$. In order to make room for $p$, GreedyLFD will evict a page of cost at most $i-1$ from its cache. As there are $k$ such pages in the cache at time $t_i$, at least one of them will not occur for the next $k-1$ requests. As there are only $k+1$ pages in total, this means that the next page fault occurs at time $t_i + k$ or later.

  In general, when the $j$th page of cost $i$ is added to GreedyLFD's cache at time $t_{i,j}$, there are $k-j+1$ pages of cost at most $i-1$ in its cache, and so the next page fault will not occur until time $t_{i,j} + k - j +1$, which gives a lower bound on $t_{i,j+1}$. Note that $t_{i,1} = t_i$

  As GreedyLFD's cost can only increase to $i+1$ once it evicts a page of cost $i$, we find that $t_{i+1}$ must occur after $k$ pages of cost $i$ have been added to its cache. GreedyLFD can choose which of the $k$ pages of cost $i$ to evict, so we find $t_{i+1} \ge t_{i,k} + k$. This yields
  \[
    t_{i+1} \ge t_{i,k} + k \ge t_{i,k-1} + k + 1 \ge t_{i,k-2} + k + 1 + 2 \ge
    \dots \ge t_{i} + k + \sum_{i=1}^k i.
  \]

  By expanding the recurrence for $i \ge 2$, we find that
  \[
  _i \ge (i-2) \left(k + \frac{k(k+1)}{2}\right) + 2k + 1.
  \]

  Using this expression, we derive an upper bound on $\cost(GreedyLFD, \sigma)$, by finding the minimum $i$ for which $\len(\sigma) \le t_i$. From our expression we find that $t_i \ge \len(\sigma)$ if $i \ge \frac{2(T - 2k - 1)}{2k + k(k+1)} + 2$, and so
  \[
    \cost(GreedyLFD, \sigma) \le \frac{2(\len(\sigma)-2k-1)}{2k + k(k+1)}+2.
  \]
\end{proof}

\begin{lemma}
  Any deterministic algorithm ALG for min-max paging with cache size $k$ is at least $\frac{k}{2}$-competitive.
\end{lemma}
\begin{proof}
  Let $n = k+1$, that is $P = \{p_1, p_2, \dots, p_{k+1}\}$. We initialize ALG and GreedyLFD with empty caches. Once ALG's cache is full, at any time step $t$, there is always one page that is not present in the cache. The adversary's strategy is always to request this page. We call this the \emph{cruel} strategy. Since ALG is deterministic, the adversary always knows which page ALG will evict from its cache if a page fault occurs, so such a sequence must always exist.

  \begin{algorithm}
    \caption{An algorithm to generate an adversarial sequence for ALG of length
      $T \ge k+1$, on which ALG faults $T$ times.}\label{cruel}
    \begin{algorithmic}[1]
      \State \Output pages $p_1, p_2, \dots, p_{k+1}$ \State $i \gets k+1$
      \While{$i < T$} \State $p \gets $ the page $p_i$ which is currently not in
      the cache of ALG; \State \Output $p$ \State $i \gets i+1$ \EndWhile
    \end{algorithmic}
  \end{algorithm}

  Let $\sigma$ be a sequence of length $T$ generated by the cruel strategy of Algorithm~\ref{cruel}.  We note that $\sigma$ causes a page fault at every step, so the total number of page faults incurred by ALG will be $T$. By a simple averaging argument, there must be at least one page that has incurred $\frac{T}{k+1}$ page faults and so $\frac{T}{k+1} \le cost(ALG, \sigma)$. On the other hand, by Lemma \ref{lfdlb}, GreedyLFD will incur a cost of at most $\frac{2(T-2k-1)}{2k+k(k+1)}+2$ while processing $\sigma$. This immediately yields
  \begin{equation*}
    \cost(\OPT, \sigma) \le \cost(GreedyLFD, \sigma) \le \frac{2(T-2k-2)}{2k + k(k+1)} + 2 \le \frac{2T}{k(k+1)} + 2\le \frac{2}{k} \cost(ALG, \sigma) + 2,
  \end{equation*}
  and so the competitive ratio is $\frac{cost(ALG, \sigma)}{\cost(OPT, \sigma)} \ge \frac{k}{2}$, since the constant vanishes as the cost grows large.
\end{proof}

Finally, we strengthen the lower bound to $\Omega(\frac{k}{\log k} \log n)$, introducing a dependence on the number of pages. We do this using the strategy presented in Algorithm~\ref{detlbnew}.

\begin{algorithm}
  \caption{An adversarial strategy for min-max paging}\label{detlbnew}
  \begin{algorithmic}[1]
    \State Let $L_m=\{p^m_0,\ldots,p^m_{n-1}\}$ be a set of $n=(k+1)^m$ pages.
    \For{$\ell$ $=$ $m$ to $1$} \State $L_{\ell-1}\gets\emptyset$.  \For{$i$ $=$
      $0$ to $(k+1)^{\ell-1}-1$} \State Use the cruel strategy of Algorithm
    \ref{cruel} on the set of $k+1$ pages
    $\{p^\ell_{(k+1)i}, p^\ell_{(k+1)i+1}, \dots, p^\ell_{(k+1)i+k} \}$ until one
    page, say $p'$, has incurred $N$ page faults since the start of this loop.
    \State $p^{\ell-1}_i$ $\gets$ $p'$ \State
    $L_{\ell-1} \gets L_{\ell-1} \cup \{ p^{\ell-1}_i \}$ \EndFor \EndFor
  \end{algorithmic}
\end{algorithm}

Intuitively, our strategy consists of splitting the $n = (k+1)^m$ pages into $(k+1)^{m-1}$ disjoint sets of $k+1$ variables. We then present the algorithm ALG with a cruel sequence for each set until one of the pages reaches cost $N$. We repeat this process layer by layer until we obtain one final page. ALG will have faulted $mN$ times on this page, while OPT will have faulted no more than roughly $\frac{N}{k} + m$ times on any page.

\detlb*
\begin{proof}
  We use the strategy defined in Algorithm \ref{detlbnew} to generate our request sequence. We observe that at the beginning of iteration $\ell$ of the outer for-loop in Algorithm \ref{detlbnew}, ALG will have faulted $(m-\ell)N$ times on each page in $\{p^{\ell}_0, \dots, p^{\ell}_{(k+1)^{\ell}-1}\}$, because we only add a page to the next level once it has incurred $N$ faults during the current level. Hence, once ALG has processed the complete sequence provided by Algorithm \ref{detlbnew}, it has cost $mN$, witnessed by page $p_0^0$.

  On the other hand, while processing the $i$th set of variables in the inner loop, the optimal offline algorithm OPT will keep page $p'$ in its cache. When $p'$ is requested for the first time in this iteration of the loop, OPT will fault once on $p'$. Afterward, $p'$ will remain in the cache of OPT until the current iteration of the inner loop finishes, incurring no more cost.

  This shows that OPT will have cost $m - \ell$ for each page $p_0^\ell, \dots, p_{(k+1)^{\ell}- 1}^\ell$ at the beginning of the $\ell$th iteration of the outer loop.

  During an iteration of the inner loop, we use the remaining $k-1$ slots in OPT's cache, which are not occupied by $p'$, to process the remaining $k$ pages in each iteration. As the cruel sequence causes ALG to fault on every request and we end it as soon as one page has faulted $N$ times, each of the remaining pages may be requested $N-1$ times. It follows that the sub-sequence $\sigma'$ of the cruel sequence, defined on the remaining $k$ pages, is of length at most $k(N-1)$. By Lemma \ref{lfdlb}, we get $\cost(\OPT, \sigma') \le \frac{2(k(N-1) - 2(k-1) - 1)}{2(k-1) + (k-1)k}+2 \le \frac{2(N-1)}{k-1} + 2$.

  Thus we find that for any page $p$, the total cost consists of the level it is raised to plus the cost incurred while processing $\sigma'$ on its last level and thus $\cost(OPT, \sigma, p, T) \le m + \frac{2(N-1)}{k-1}+2$ and so
  \[
    \frac{\cost(ALG, \rho)}{\cost(OPT,\rho)} \ge \frac{mN}{m + 2 +
      \frac{2(N-1)}{k-1}} = \frac{(k-1)mN}{(k-1)(m+2) + 2(N-1)}.
  \]
  As $N$ grows large, the right hand side will converge to $\frac{(k-1)m}{2} = \frac{k-1}{2} \log_{k+1} n$.
\end{proof}




\section{Deferred Proofs}

This section contains some deferred proofs from the paper.

\fenchelmonotone*
\begin{proof}
Indeed, let $c \in \R_{+}^T$, then 
\[
f^*(y) = \sup_{w \in \R_{+}^T} \langle y, w \rangle - f(w) \le \sup_{w
\in \R_{+}^T} \langle y, w \rangle + \langle c, w \rangle - f(w) = f^*(y +
c),
\]
as $\langle c, w\rangle$ is always non-negative.
\end{proof}

\weakduality*
\begin{proof} 
Let $b = \begin{bmatrix} B(1)-k \\ \vdots \\ B(T) - k \end{bmatrix}$. As $x$ is feasible, it must satisfy $A x \ge b$, which gives $b - Ax \le \zeros$, and similarly from $x \le \ones$, we get $x - \ones \le \zeros$. As all entries are negative, taking the inner product of these vectors with the non-negative vectors $y$ and $z$, respectively, will yield a negative number. Hence, we get the first inequality in the following chain of inequalities:
\begin{align*}
  f(x) &\ge f(x) + y^\top ((|B(t)|-k) \ones - Ax) + z^\top (x -\ones) \\
       &= \sum_{t = 1}^{T} (|B(t)|-k)y_t - \sum_{p=1}^{n} \sum_{j=1}^{r(p,T)} z_{p,j} - (y^\top Ax -z^\top x - f(x))\\
       &= \sum_{t=1}^{T} (|B(t)|-k)y_t -  \sum_{p=1}^{n} \sum_{j=1}^{r(p,T)} z_{p,j} -(\langle A^\top y - z, x\rangle - f(x))\\
       &\ge \sum_{t=1}^{T} (|B(t)|-k)y_t -  \sum_{p=1}^{n} \sum_{j=1}^{r(p,T)} z_{p,j} - \sup_{w \in \R_+^T} (\langle A^\top y - z, w\rangle -f(w))\\
       &= \sum_{t=1}^{T} (|B(t)|-k)y_t -  \sum_{p=1}^{n} \sum_{j=1}^{r(p,T)} z_{p,j} - f^*(A^\top y -z).\\
\end{align*}
All equalities are obtained via simple rearranging of terms, and the final inequality is due to the definition of the supremum.
\end{proof}

\xlowerbound*
\begin{proof}
First, observe that $x_{p,j}$ starts increasing at time $\tau(t(p,j))$, the time at which we finish processing the $j$'th request to $p$. $x_{p,j}$ keeps increasing until one of the two events happens: $x_{p,j}$ reaches $1$, or we get the next request to $p$, after which it remains constant till the end. Let $\tau^{\text{end}}$ denote the time at which either of these events happens.

For $\tau\in[\tau(t(p,j)),\tau^{\text{end}})$, we have
\[\frac{dx_{p,j}(\tau)}{d\tau}=s_{p,j}(\tau)\cdot\left(x_{p,j}(\tau)+\frac{1}{k}\right)\geq s_{p,j}'\cdot\left(x_{p,j}(\tau)+\frac{1}{k}\right)\text{,}\]
and therefore,
\[\frac{1}{s_{p,j}'}\frac{d}{d\tau}\ln\left(x_{p,j}(\tau)+\frac{1}{k}\right)=\frac{1}{s_{p,j}'\cdot(x_{p,j}(\tau)+1/k)}\cdot\frac{dx_{p,j}(\tau)}{d\tau}\geq1\text{.}\]
Integrating over the interval $[\tau(t(p,j)),\tau^{\text{end}})$, we get,
\begin{equation}\label{eqn_x}
\frac{1}{s_{p,j}'}\ln\left(\frac{\bar{x}_{p,j}+1/k}{1/k}\right)\geq\tau^{\text{end}}-\tau(t(p,j))\text{.}
\end{equation}

Next, observe that the variable $y_t$ starts increasing from $0$ at the uniform rate $r$ at time $\tau(t-1)$ and stops increasing at time $\tau(t)$. Thus, $\bar{y}_t=r\cdot(\tau(t)-\tau(t-1))$. Summing over all $t$ from $t(p,j)+1$ to $t(p,j+1)-1$, we get,
\begin{equation}\label{eqn_y}
\frac{1}{r}\cdot\left(\sum_{t=t(p,j)+1}^{t(p,j+1)-1}\bar{y}_t\right)=\tau(t(p,j+1)-1)-\tau(t(p,j))\text{.}
\end{equation}

Finally, consider the variable $z_{p,j}$. If $x_{p,j}$ stops increasing because the next request to $p$ arrives, then $\bar{z}_{p,j}=0$ and $\tau^{\text{end}}=\tau(t(p,j+1)-1)$. Using Equation (\ref{eqn_y}), we get,
\[
\frac{1}{r}\cdot\left(\sum_{t=t(p,j)+1}^{t(p,j+1)-1}\bar{y}_t-\bar{z}_{p,j}\right)=\tau(t(p,j+1)-1)-\tau(t(p,j))=\tau^{\text{end}}-\tau(t(p,j))\text{.}
\]
On the other hand, if $x_{p,j}$ stops increasing because it reaches $1$, then $z_{p,j}$ starts increasing from $0$ at the uniform rate $r$ at time $\tau^{\text{end}}$, and stops increasing at time $\tau(t(p,j+1)-1)$. Thus, $\bar{z}_{p,j}=r\cdot(\tau(t(p,j+1)-1)-\tau^{\text{end}})$. Again, using Equation (\ref{eqn_y}), we get,
\[
\frac{1}{r}\cdot\left(\sum_{t=t(p,j)+1}^{t(p,j+1)-1}\bar{y}_t-\bar{z}_{p,j}\right)=(\tau(t(p,j+1)-1)-\tau(t(p,j)))-(\tau(t(p,j+1)-1)-\tau^{\text{end}})=t^{\text{end}}-\tau(t(p,j))\text{.}
\]
Thus, in either case, we have,
\begin{equation}\label{eqn_yz}
\frac{1}{r}\cdot\left(\sum_{t=t(p,j)+1}^{t(p,j+1)-1}\bar{y}_t-\bar{z}_{p,j}\right)=\tau^{\text{end}}-\tau(t(p,j))\text{.}
\end{equation}

From Inequality (\ref{eqn_x}) and Equation (\ref{eqn_yz}), we get,
\[
\frac{1}{s_{p,j}'}\ln\left(\frac{\bar{x}_{p,j}+1/k}{1/k}\right)\geq\tau^{\text{end}}-\tau(t(p,j))=\frac{1}{r}\cdot\left(\sum_{t=t(p,j)+1}^{t(p,j+1)-1}\bar{y}_t-\bar{z}_{p,j}\right)\text{.}
\]
Rearranging, we get,
\[    \bar{x}_{p,j} \ge \frac1k \left(\exp \left(
        \frac{ s_{p,j}'}{r} \left( \sum_{t = t(p, j)+1}^{t(p,j+1)-1} \bar{y}_t -
          \bar{z}_{p,j} \right) \right) - 1\right),
\]
as required.
\end{proof}

\feasibility*
\begin{proof}
The first statement follows immediately from the definition of the algorithm and the fact that we can always fulfill the primal constraints, for example, by setting all the variables to $1$.

To show the second statement, we note that $\bar{x} \le \mathbf{1}$, which with Equation \eqref{lem2eq} gives us
\[ 
\frac1k \left(\exp \left( \frac{ s_{p,j}'}{r} \left( \sum_{t = t(p,
j)+1}^{t(p,j+1)-1} \bar{y}_t - \bar{z}_{p,j} \right) \right) - 1\right) \le
\bar{x}_{p,j} \le 1.
\] 
Taking the left- and right-hand sides gives us, after rearranging and taking
logarithms,
\[
\sum_{t = t(p, j)+1}^{t(p,j+1)-1} \bar{y}_t - \bar{z}_{p,j} \le
\frac{r}{s_{p,j}'} \ln(k + 1),
\] 
where the left hand side is $(A^T \bar{y} - \bar{z})_{p,j}$ as a $1$ only appears in the $(p,j)$th column of $A$ from row $t(p,j)+1$ through row $t(p,j+1)-1$.
\end{proof}


\roundinghelper*
\begin{proof}
  We define our rounded solution $x^{*}$ as
  \begin{equation*}
    x_{p,j}^{*} = \begin{cases} 0 & \text{if $x_{p,j} < \frac{1}{8k}$,} \\ \min\{\frac{1}{4k} \lceil 8k x_{p,j} \rceil, 1\} & \text{otherwise.} \end{cases}
  \end{equation*}
  That is, we round every variable $x_{p,j}$ of value less than $\frac{1}{8k}$ to $0$, and every variable of greater value will be doubled and then rounded up to the nearest multiple of $\frac{1}{4k}$.

  As each variable's value is at most doubled and then rounded up, the fractional cost for min-max paging can at most triple, as each variable in $x^{*}$ is no larger than $3$ times the corresponding variable in $x$ and the objective function $f(x)$ in min-max paging satisfies $f(cx) = cf(x)$ for any $c \in \mathbb{R}$. It remains to show that the solution $x^{*}$ is feasible.

  We note that the only variables whose value in $x^{*}$ can decrease during the rounding are those of value less than $\frac{1}{8k}$. The total number of such variables other than $p_{t}$ in a feasible solution is at most $k$, as otherwise.
  \[
    \sum_{B(t) \setminus \{p_{t}\}} x_{p,r(p,t)} \le n - 1 - k + \frac{1}{8k} k \le n-k.
  \]

  Hence we note that the total contribution $L$ of these variables to the constraint of round $t$ is at most $\frac{1}{8}$.

  As $n - k$ is an integer, and the fractional algorithm stops ejecting pages as soon as the constraint is satisfied, we know that $\sum_{p \in B(t) \setminus \{p_{t}\}} x_{p,r(p,t)}$ is integral.

  We further note that the contribution of all variables that are rounded up in this bound must therefore be at least $n - k - \frac18$

  If any variable $x_{p,j}$ in $x$ fulfills $\frac18 \le x_{p,j} \le \frac78$, then the rounding step will increase this variable by at least $\frac18$, which compensates the value lost by rounding down all small variables.

  Otherwise, if we let $A = \{x_{p,r(p,t)} \in B(t) \setminus \{p_{t}\} \mid x_{p,r(p,t)} \le \frac18 \}$ and $\sum_{x \in A} x \ge \frac18$, then the doubling of these variables compensates for the rounding down of small variables.

  Finally, if neither of these is the case, the variables of size greater than $\frac78$ must sum up to at most $n - k - L$, where $L$ denotes the amount of mass lost by rounding the small variables to $0$,  and at least $n - k - 2L > n - k - 1$. The sum of the rounded-up variables is an integer, as all variables of value greater than $\frac12$ are rounded to $1$, so we gain at least $L$ from rounding up the large variables.
\end{proof}

\begin{figure}
    \centering
    \[
    \begin{blockarray}{cccccccccccc}
     & \dots & x_{p_1,7} & x_{p_2,9} & x_{p_3, 3} & x_{p_1,8} & x_{p_4,5} & x_{p_2,10} & x_{p_5,3} & x_{p_1, 9} & \dots \\
    \begin{block}{c[ccccccccccc]}
    \vdots & \ddots & \vdots & \vdots & \vdots & \vdots & \vdots & \vdots & \vdots & \vdots &  \\
    t(p_1, 7) & \dots & 0 & 0 & 0 & 0 & 0 & 0 & 0 & 0 & \dots \\
    t(p_2, 9) & \dots & 1 & 0 & 0 & 0 & 0 & 0 & 0 & 0 & \dots \\
    t(p_3, 3) & \dots & 1 & 1 & 0 & 0 & 0 & 0 & 0 & 0 & \dots \\
    t(p_1, 8) & \dots & 0 & 1 & 1 & 0 & 0 & 0 & 0 & 0 & \dots \\
    t(p_4, 5) & \dots & 0 & 1 & 1 & 1 & 0 & 0 & 0 & 0 & \dots \\
    t(p_2, 10) & \dots & 0 & 0 & 1 & 1 & 1 & 0 & 0 & 0 & \dots \\
    t(p_5, 3) & \dots & 0 & 0 & 1 & 1 & 1 & 1 & 0 & 0 & \dots \\
    t(p_1, 9) & \dots & 0 & 0 & 1 & 0 & 1 & 1 & 1 & 0 & \dots \\
    \myvdots & \dots & 0 & 0 & 1 & 0 & 1 & 1 & 1 & 1 & \dots \\
    \vdots &  & \vdots & \vdots & \vdots & \vdots & \vdots & \vdots & \vdots & \vdots & \ddots \\
    \end{block}
    \end{blockarray}\vspace*{-1.25\baselineskip}
    \]
    \caption{An example of the structure of a constraint matrix of the paging problem, corresponding to the sub-sequence of requests $p_1, p_2, p_3, p_1, p_4, p_2, p_5, p_1$ starting at time $t' = t(p_1, 7)$. The matrix is lower triangular, due to the columns being in order of the appearance of variables. In this example, we assume that the pages $p_1, p_2, p_3, p_4$ and $p_5$ have been requested $6, 8 , 2 , 4,$ and $2$ times before the appearance of this sequence respectively. }
    
    \label{fig:matrix}
  \end{figure}
\end{document}